\documentclass[12pt]{article}

\usepackage{amssymb,amsmath}
\usepackage{epsfig}
\usepackage{epsf}

\def\BibTeX{{\rm B\kern-.05em{\sc i\kern-.025em b}\kern-.08em
    T\kern-.1667em\lower.7ex\hbox{E}\kern-.125emX}}

\def\argmax{\mathrm{argmax}}
\def\mathbb{\mathbf}
\newtheorem{theorem}{Theorem}
\newtheorem{algorithm}{Algorithm}
\newtheorem{proposition}{Proposition}
\newtheorem{lemma}{Lemma}

\newtheorem{procedure}{Procedure}

\newtheorem{assumptions}{Assumptions} 

\newcommand{\qed}{\mbox{} \hfill $\Box$ }

\title{Kullback Proximal Algorithms for Maximum Likelihood Estimation\footnotemark[1]}
\author{St\'ephane Chr\'etien, Alfred O. Hero}

\begin{document}
  
\footnotetext[1]{Stephane Chretien is with Universit\'{e} Libre de Bruxelles,
  Campus de la Plaine, CP 210-01, 1050 Bruxelles, Belgium,
  (schretie@smg.ulb.ac.be) and Alfred Hero is with the Dept.  of Electrical
  Engineering and Computer Science, 1301 Beal St., University of Michigan,
  Ann Arbor, MI 48109-2122 (hero@eecs.umich.edu).  This research was
  supported in part by AFOSR grant F49620-97-0028}

\date{}
\maketitle

\baselineskip 0.3in
\parskip=0.2in
\itemsep 0.2in

\markboth{}{\hfill Chretien, Hero, ``Kullback Proximal Algorithms
    ... (submitted to IEEE IT Trans.),'' \today}
  
  \abstract{Accelerated algorithms for maximum likelihood image
    reconstruction are essential for emerging applications such as 3D
    tomography, dynamic tomographic imaging, and other high dimensional
    inverse problems.  In this paper, we introduce and analyze a class of
    fast and stable sequential optimization methods for computing maximum
    likelihood estimates and study its convergence properties. These methods
    are based on a {\it proximal point algorithm} implemented with the
    Kullback-Liebler (KL) divergence between posterior densities of the
    complete data as a proximal penalty function.  When the proximal
    relaxation parameter is set to unity one obtains the classical
    expectation maximization (EM) algorithm.  For a decreasing sequence of
    relaxation parameters, relaxed versions of EM are obtained which can have
    much faster asymptotic convergence without sacrifice of monotonicity.  We
    present an implementation of the algorithm using Mor\'{e}'s {\it Trust
      Region} update strategy. For illustration the method is applied to a
    non-quadratic inverse problem with Poisson distributed data.}
  
\noindent{\bf Keywords}: {\it
accelerated EM algorithm, Kullback-Liebler relaxation, proximal
    point iterations, superlinear convergence, Trust Region methods, emission
    tomography.}

\newpage

\noindent{ \bf LIST OF FIGURES}

\begin{enumerate}
\item
{Two rail phantom for 1D deblurring example.}
\item
Blurred two level phantom. Blurring kernel is Gaussian with standard
  width approximately equal to rail separation distance in phantom. An
  additive randoms noise of 0.3 was added.
\item
Snapshot of log--Likelihood vs iteration for plain 
  EM and KPP EM algorithm.  Plain EM initially produces greater increases in
  likelihood function but is overtaken by KPP EM at 7 iterations and
  thereafter.
\item
The sequence $\log \|\theta_k-\theta^*\|$ vs iteration for plain EM
  and  KPP EM algorithms.  Here $\theta^*$ is limiting value for each
  of the algorithms. Note the superlinear convergence of KPP. 
\item 
Reconstructed images after 150 iterations of plain
  EM and KPP EM algorithms.
\item
Evolution of the reconstructed source vs iteration
  for plain EM and KPP EM.
\end{enumerate}

\newpage

\parskip=0.2in

\section{Introduction}

Maximum likelihood (ML) or maximum penalized likelihood (MPL) approaches have
been widely adopted for image restoration and image reconstruction from noise
contaminated data with known statistical distribution.  In many cases the
likelihood function is in a form for which analytical solution is difficult
or impossible.  When this is the case iterative solutions to the ML
reconstruction or restoration problem are of interest.  Among the most stable
iterative strategies for ML is the popular expectation maximization (EM)
algorithm \cite{Dempster&etal:JRSS77}. The EM algorithm has been widely
applied to emission and transmission computed tomography
\cite{Shepp&Vardi:MI82,Lange:84:emr,Ollinger&Snyder:NS85} with Poisson data.
The EM algorithm has the attractive property of monotonicity which guarantees
that the likelihood function increases with each iteration.  The convergence
properties of the EM algorithm and its variants have been extensively studied
in the literature; see \cite{Wu:AnnStat83} and \cite{Hero&Fessler:SS95} for
instance.  It is well known that under strong concavity assumptions the EM
algorithm converges linearly towards the ML estimator $\theta_{ML}$. However,
the rate coefficient is small and in practice the EM algorithm suffers from
slow convergence in late iterations.  Efforts to improve on the asymptotic
convergence rate of the EM algorithm have included: Aitken's acceleration
\cite{Louis:JRSS82}, over-relaxation \cite{Lewitt&Muehllehner:MI86},
conjugate gradient \cite{Kaufman:MI87} \cite{Jamshidian&Jennrich:JASA93},
Newton methods \cite{Meilijson:JRSS89} \cite{Bouman&Sauer:ISS93},
quasi-Newton methods \cite{Lange:SS95}, ordered subsets EM
\cite{Hudson&Larkin:MI94} and stochastic EM \cite{Lavielle:SP95}.
Unfortunately, these methods do not automatically guarantee the monotone
increasing likelihood property as does standard EM. Furthermore, many of
these accelerated algorithms require additional monitoring for instability
\cite{Lansky&Casella:CSS90}.  This is especially problematic for high
dimensional image reconstruction problems, e.g. 3D or dynamic imaging, where
monitoring could add significant computational overhead to the reconstruction
algorithm.

The contribution of this paper is the introduction of a class of accelerated
EM algorithms for likelihood function maximization via exploitation of a
general relation between EM and proximal point (PP) algorithms. These
algorithms converge and can have quadratic rates of convergence even with
approximate updating. Proximal point algorithms were introduced by Martinet
\cite{Martinet:1970} and Rockafellar \cite{Rockafellar:1976}, based on the
work of Minty \cite{Minty:Duke62} and Moreau \cite{Moreau:BSMF65}, for the
purpose of solving convex minimization problems with convex constraints.  A
key motivation for the PP algorithm is that by adding a sequence of
iteration-dependent penalties, called proximal penalties, to the objective
function to be maximized one obtains stable iterative algorithms which
frequently outperform standard optimization methods without proximal
penalties, e.g. see Goldstein and Russak \cite{Goldstein&Russak:NFAO87}.
Furthermore, the PP algorithm plays a paramount role in non-differentiable
optimization due to its connections with the Moreau-Yosida regularization;
see Minty \cite{Minty:Duke62}, Moreau \cite{Moreau:BSMF65}, Rockafellar
\cite{Rockafellar:1976} and Hiriart-Hurruty and Lemar\'echal \cite{Hiriart&Lemarechal:93}.

While the original PP algorithm used a simple quadratic penalty more general
versions of PP have recently been proposed which use non-quadratic penalties,
and in particular entropic penalties.  Such penalties are most commonly
applied to ensure non-negativity when solving Lagrange duals of inequality
constrained primal problems; see for example papers by Censor and Zenios
\cite{Censor&Zenios:JOTA92}, Ekstein \cite{Ekstein:MOR93}, Eggermont
\cite{Eggermont:1990}, and Teboulle \cite{Teboulle:1992}.  In this paper we
show that by choosing the proximal penalty function of PP as the
Kullback-Liebler (KL) divergence between successive iterates of the posterior
densities of the complete data, a generalization of the generic EM maximum
likelihood algorithm is obtained with accelerated convergence rate.  When the
relaxation sequence is constant and equal to unity the PP algorithm with KL
proximal penalty reduces to the standard EM algorithm. On the other hand for
a decreasing relaxation sequence the PP algorithm with KL proximal penalty is
shown to yield an iterative ML algorithm which has much faster convergence
than EM without sacrificing its monotonic likelihood property.
  
It is important to point out that relations between particular EM and
particular PP algorithms have been previously observed, but not in the full
generality established in this paper.  Specifically, for parameters
constrained to the non-negative orthant, Eggermont \cite{Eggermont:1990}
established a relation between an entropic modification of the standard PP
algorithm and a class of multiplicative methods for smooth convex
optimization. The modified PP algorithm that was introduced in
\cite{Eggermont:1990} was obtained by replacing the standard quadratic
penalty by the relative entropy between {\it successive non-negative
  parameter iterates}.  This extension was shown to be equivalent to an
``implicit'' algorithm which, after some approximations to the exact PP
objective function, reduces to the ``explicit'' Shepp and Vardi EM algorithm
\cite{Shepp&Vardi:MI82} for image reconstruction in emission tomography.
Eggermont \cite{Eggermont:1990} went on to prove that the explicit and
implicit algorithms are monotonic and both converge when the sequence of
relaxation parameters is bounded below by a strictly positive number.

In contrast to \cite{Eggermont:1990}, here we establish a general and exact
relation between the generic EM procedure, i.e. arbitrary incomplete and
complete data distributions, and an extended class of PP algorithms. As
pointed out above, the extended PP algorithm is implemented with a proximal
penalty which is the relative entropy (KL divergence) between {\it successive
  iterates of the posterior densities of the complete data}. This
modification produces a class of algorithms which we refer to as
Kullback-Liebler proximal point (KPP).  We prove a global convergence result
for the KPP algorithm under strict concavity assumptions.  An approximate KPP
is also proposed using the Trust Region strategy \cite{More:83,Nocedal:1999}
adapted to KPP.  We show, in particular, that both the exact and approximate
KPP algorithms have superlinear convergence rates when the sequence of
positive relaxation parameters converge to zero.  Finally, we illustrate
these results for KPP acceleration of the Shepp and Vardi EM algorithm
implemented with Trust Region updating.

The results given here are also applicable to the non-linear updating methods
of Kivinen and Warmuth \cite{Kivinen&Warmuth:InfoComp97} for accelerating the
convergence of Gaussian mixture-model identification algorithms in supervised
machine learning, see also Warmuth and Azoury \cite{Warmuth&Azoury:UAI99} and
Helmbold, Schapire, Singer and Warmuth \cite{Helmbold&etal:JML97}. Indeed,
similarly to the general KPP algorithm introduced in this paper, in
\cite{Helmbold&etal:JML97} the KL divergence between the new and the old
mixture model was added to the gradient of the Gaussian mixture-model
likelihood function, appropriately weighted with a multiplicative factor
called the learning rate parameter. This procedure led to what the authors of
\cite{Helmbold&etal:JML97} called an exponentiated gradient algorithm.  These
authors provided experimental evidence of significant improvements in
convergence rate as compared to gradient descent and ordinary EM. The results
in this paper provide a general theory which validate such experimental
results for a very broad class of parametric estimation problems.

The outline of the paper is as follows. In Section \ref{sec:background} we
provide a brief review of key elements of the classical EM algorithm. In
Section \ref{proxEM}, we establish the general relationship between the EM
algorithm and the proximal point algorithm. In section \ref{Kullprox}, we
present the general KPP algorithm and we establish global and superlinear
convergence to the maximum likelihood estimator for a smooth and strictly
concave likelihood function. In section \ref{approx}, we study second order
approximations of the KPP iteration using Trust Region updating. Finally, in
Section \ref{sec:numerical} we present numerical comparisons for a Poisson
inverse problem.

\section{Background}
\label{sec:background}

The problem of maximum likelihood (ML) estimation consists of finding
a solution of the form
\begin{equation}
\label{ml}
\theta_{ML} = {\rm argmax}_{\theta \in {\mathbb R}^p} \; l_y(\theta), 
\end{equation}
where $y$ is an observed sample of a random variable $Y$ defined on a
sample space $\mathcal Y$ and $l_y(\theta)$ is the log-likelihood
function defined by
\begin{equation}
l_y(\theta)=\log g(y;\theta),
\end{equation}
and $g(y;\theta)$ denotes the density of $Y$ at $y$ parametrized by a vector
parameter $\theta$ in ${\mathbb R}^p$. One of the most popular iterative
methods for solving ML estimation problems is the Expectation Maximization
(EM) algorithm described in Dempster, Laird, and Rubin
\cite{Dempster&etal:JRSS77} which we recall for the reader.

A more informative data space $\mathcal X$ is introduced. A random variable
$X$ is defined on $\mathcal X$ with density $f(x;\theta)$ parametrized by
$\theta$.  The data $X$ is more informative than the actual data $Y$ in the
sense that $Y$ is a compression of $X$, i.e. there exists a non-invertible
transformation $h$ such that $Y=h(X)$.  If one had access to the data $X$ it
would therefore be advantageous to replace the ML estimation problem
(\ref{ml}) by
\begin{equation}
\label{mlx}
\hat{\theta}_{ML}=\argmax_{\theta \in {\mathbb R}^p} l_x(\theta), 
\end{equation}
with $l_x(\theta)=\log f(x;\theta)$. 
Since $y=h(x)$ the density $g$ of $Y$ is related to the density $f$ of $X$
through 
\begin{equation}
\label{cond}
g(y;\theta)=\int_{h^{-1}(\{y\})}f(x;\theta)d\mu(x) 
\end{equation}
for an appropriate measure $\mu$ on $\mathcal X$.  
In this setting, the data $y$ are called {\em
incomplete data} whereas the data $x$ are called {\em complete data}.

Of course the complete data $x$ corresponding to a given observed sample $y$
are unknown. Therefore, the complete data likelihood function $l_x(\theta)$
can only be estimated. Given the observed data $y$ and a previous estimate of
$\theta$ denoted $\bar{\theta}$, the following minimum mean square error
estimator (MMSE) of the quantity $l_x(\theta)$ is natural
\begin{equation*}
Q(\theta,\bar{\theta})={\sf E}[\log f(x;\theta)| y;\bar{\theta}],
\end{equation*} 
where, for any integrable function $F(x)$ on $\mathcal X$, we have
defined the conditional expectation
\begin{equation*}
{\sf E}[F(x)| y;\bar{\theta}]=\int_{h^{-1}(\{y\})}
F(x) k(x| y;\bar{\theta}) d\mu(x)
\end{equation*}
and $k(x| y;\bar{\theta})$ is the conditional density function given $y$
\begin{equation}\label{condi}
k(x| y;\bar{\theta})=\frac{f(x;\bar{\theta})}{g(y;\bar{\theta})}.
\end{equation}
The EM algorithm generates a sequence of approximations to the solution
(\ref{mlx}) starting from an initial guess $\theta^0$ of $\theta_{ML}$
and is defined by
\begin{equation*}
\text{\bf Compute } Q(\theta,\theta^k)={\sf E}[\log f(x;\theta)| y;\theta^k] 
\text{\hspace{1cm}\bf E Step}\nonumber 
\end{equation*}
\begin{equation*}
\theta^{k+1}={\rm argmax}_{\theta \in {\mathbb R}^p} Q(\theta,\theta^k)
\text{\hspace{2.9cm}\bf M Step}\nonumber
\end{equation*}

A key to understanding the convergence of the EM algorithm is the
decomposition of the likelihood function presented in Dempster, Laird and
Rubin \cite{Dempster&etal:JRSS77}.  As this decomposition is also the prime
motivation for the KPP generalization of EM it will be worthwhile to recall
certain elements of their argument.  The likelihood can be decomposed as
\begin{equation}
l_y(\theta)=Q(\theta,\bar{\theta})+H(\theta,\bar{\theta})
\label{eq:decomp}
\end{equation}
where 
\begin{equation*}
H(\theta,\bar{\theta})=-{\sf E}[ \log k(x| y;\theta)| y;\bar{\theta}].
\end{equation*}
It follows from elementary application of Jensen's
inequality to the log function that
\begin{equation}
\label{entrop}
H(\theta,\bar{\theta})\geq H(\theta,\theta)\geq 0,\; \:\: \forall
\theta,\;\bar{\theta} \in {\mathbb R}^p.
\end{equation}

Observe from (\ref{eq:decomp}) and (\ref{entrop}) that for any $\theta^{k}$
the $\theta$ function $Q(\theta,\theta^{k})$ is a lower bound on the log
likelihood function $l_y(\theta)$. This property is sufficient to ensure
monotonicity of the algorithm. Specifically, since the the M-step implies that
\begin{equation}
\label{incrQ}
Q(\theta^{k+1},\theta^k)\geq Q(\theta^k,\theta^k),
\end{equation}
one obtains
\begin{eqnarray}
l_y(\theta^{k+1})-l_y(\theta^k)&\geq & Q(\theta^{k+1},\theta^k)
-Q(\theta^k,\theta^k)\label{eq:monot}\\
&& +H(\theta^{k+1},\theta^k)-H(\theta^k,\theta^k). 
\nonumber \end{eqnarray}
Hence, using (\ref{incrQ}) and (\ref{entrop})
\begin{equation*}
l_y(\theta^{k+1})\geq l_y(\theta^k).
\end{equation*}
This is the well known monotonicity property of the EM algorithm.

Note that if the function $H(\theta,\bar{\theta}) $ in (\ref{eq:decomp}) were
scaled by an arbitrary positive factor $\beta$ the function
$Q(\theta,\bar{\theta})$ would remain a lower bound on $l_y(\theta)$, the
right hand side of (\ref{eq:monot}) would remain positive and monotonicity of
the algorithm would be preserved. As will be shown below, if $\beta$ is
allowed to vary with iteration in a suitable manner one obtains a monotone,
superlinearly convergent generalization of the EM algorithm.

\section{Proximal point methods and the EM algorithm}
\label{proxEM}
In this section, we present the proximal point (PP) algorithm of Rockafellar
and Martinet. We then demonstrate that EM is a particular case of proximal
point implemented with a Kullback-type proximal penalty.

\subsection{The proximal point algorithm}
Consider the general problem of maximizing a concave function $\Phi(\theta)$.
The proximal point algorithm is an iterative procedure which can be
written
\begin{equation}
\label{proxit}
\theta^{k+1}={\rm argmax}_{\theta \in {\mathbb R}^p}\left\{\Phi(\theta)
-\frac{\beta_k}{2} \|\theta-\theta^k\|^2\right\}.
\end{equation}
The quadratic penalty $\|\theta-\theta^k\|^2$ is relaxed using a
sequence of positive parameters $\{\beta_k\}$. In
\cite{Rockafellar:1976}, Rockafellar showed that superlinear
convergence of this method is obtained when the sequence $\{\beta_k\}$
converges towards zero.  In numerical implementations of proximal
point the function $\Phi(\theta)$ is generally replaced by a piecewise
linear model \cite{Hiriart&Lemarechal:93}.

\subsection{Proximal interpretation of the EM algorithm}
In this section, we establish an exact relationship between the
generic EM procedure and an extended proximal point algorithm.  For our
purposes, we will need to consider a particular Kullback-Liebler (KL)
information measure.  Assume that the family of conditional densities $\{k(x|
y;\theta)\}_{\theta \in {\mathbb R}^p}$ is regular in the sense of Ibragimov
and Khasminskii \cite{Ibragimov&Has'minskii:81}, in particular $k(x|
y;\theta)\mu(x)$ and $k(x| y;\bar{\theta)}\mu(x)$ are mutually absolutely
continuous for any $\theta$ and $\bar{\theta}$ in ${\mathbb R}^p$. Then the
Radon-Nikodym derivative $\frac{k(x| y,\bar{\theta})}{k(x| y;\theta)}$ exists
for all $\theta,\bar{\theta}$ and we can define the following KL divergence:
\begin{equation}
\label{kullb}
I_y(\bar{\theta},\theta)={\sf E}\bigl[
\log \frac{k(x| y,\bar{\theta})}{k(x| y;\theta)}| y;\bar{\theta} \; 
\bigr].
\end{equation}
\begin{proposition}
\label{equiem}
The EM algorithm is equivalent to the following recursion with $\beta_k =1$,
$k=1, 2,\ldots, $
\begin{equation}
\theta^{k+1}={\rm argmax}_{\theta\in {\mathbb R}^p} \left\{
l_y(\theta)- \beta_k I_y(\theta^k,\theta)\right\}
\label{iteration}
\end{equation}
\end{proposition}

For general positive sequence $\{\beta_k\}$ the recursion in Proposition
\ref{equiem} can be identified as a modification of the PP algorithm
(\ref{proxit}) with the standard quadratic penalty replaced by the KL penalty
(\ref{kullb}) and having relaxation sequence $\{\beta_k\}$.  In the sequel we
call this modified PP algorithm the Kullback-Liebler proximal point (KPP)
algorithm.  In many treatments of the EM algorithm the quantity 
$$
Q(\theta, \bar{\theta}) = l_y(\theta) - l_y(\bar{\theta}) -
I(\bar{\theta},\theta) 
$$
is the surrogate function that is maximized in the M-step. This surrogate
objective function is identical (up to an additive constant) to the KPP
objective $l_y(\theta)- \beta_k I_y(\theta^k,\theta)$ of (\ref{iteration})
when $\beta_k=1$.

\noindent{\it Proof of Proposition \ref{equiem}:}
The key to making the connection with the proximal point algorithm
is the following representation of the M step:
\begin{equation}
\theta^{k+1}={\rm argmax}_{\theta\in {\mathbb R}^p}
\bigl\{\log g(y;\theta)+{\sf E}\bigl[\log \frac{f(x;\theta)}
{g(y;\theta)} | y;\theta^k\bigr]\bigr\}.
\nonumber
\end{equation}
This equation is equivalent to
\begin{align}
\theta^{k+1}= {\rm argmax}_{\theta \in {\mathbb R}^p} & \bigl\{
\log g(y;\theta) +{\sf E}\bigl[\log \frac{f(x;\theta)}
{g(y;\theta)} | y;\theta^k\bigr] \nonumber \\
& -{\sf E}\bigl[\log \frac{f(x;\theta^k)}
{g(y;\theta^k)} | y;\theta^k\bigr]\bigr\}\nonumber
\end{align}
since the additional term is constant in $\theta$.
Recalling that $k(x| y;\theta)=\frac{f(x;\theta)}
{g(y;\theta)}$,
\begin{align}
\theta^{k+1}= {\rm argmax}_{\theta \in {\mathbb R}^p} & \bigl\{
\log g(y;\theta)+{\sf E}\bigl[\log k(x| y;\theta) | y;\theta^k\bigr] \nonumber\\
& -{\sf E}\bigl[\log k(x| y;\theta^k) | y;\theta^k\bigr] \bigr\}.\nonumber
\end{align}
We finally obtain
\begin{equation}
\theta^{k+1}={\rm argmax}_{\theta \in {\mathbb R}^p}\bigl\{
\log g(y;\theta)+{\sf E}\bigl[\log \frac{k(x| y;\theta)}{k(x| y;\theta^k)} 
| y;\theta^k\bigr]\bigr\}\nonumber 
\end{equation}
which concludes the proof.  \qed

\section{Convergence of the KPP Algorithm}
\label{Kullprox}

In this section we establish monotonicity and other convergence properties of
the KPP algorithm of Proposition \ref{equiem}.

\subsection{Monotonicity}
For bounded domain of $\theta$, the KPP algorithm is well defined since the
maximum in (\ref{iteration}) is always achieved in a bounded set.
Monotonicity is guaranteed by this procedure as proved in the following
proposition.
\begin{proposition}
\label{inc}
The log-likelihood sequence $\{l_y(\theta^k)\}$ is monotone
non-decreasing and satisfies
\begin{equation}
\label{gultov}
l_y(\theta^{k+1})-l_y(\theta^k)\geq \beta_k I_y(\theta^k,\theta^{k+1}) ,
\end{equation}
\end{proposition}

\noindent{ \it Proof:}
From the recurrence in (\ref{iteration}), we have
\begin{equation*}
l_y(\theta^{k+1})-l_y(\theta^k)\geq \beta_k I_y(\theta^k,\theta^{k+1})-
\beta_k I_y(\theta^k,\theta^k). 
\end{equation*}
Since $I_y(\theta^k,\theta^k)=0$ and $I_y(\theta^k,\theta^{k+1})\geq 0$, we 
deduce (\ref{gultov}) and that $\{l_y(\theta^k)\}$ is 
non-decreasing. \qed

We next turn to asymptotic convergence of the KPP iterates $\{\theta^k\}$.

\subsection{Asymptotic Convergence}

In the sequel $\nabla_{01} I_y(\bar{\theta},\theta)$ (respectively
$\nabla_{01}^2 I_y(\bar{\theta},\theta)$) denotes the gradient (respectively
the Hessian matrix) of $I_y(\bar{\theta},\theta)$ in the first variable.  For
a square matrix $M$, $\Lambda_M$ denotes the greatest eigenvalue of a matrix
$M$ and $\lambda_M$ denotes the smallest.

We make the following assumptions

\begin{assumptions}
\label{ass}
We assume the following:
\begin{itemize}
\item[(i)] $l_y(\theta)$ is twice continuously differentiable on
${\mathbb R}^p$
and $I_y(\bar{\theta},\theta)$ is twice continuously differentiable in
$(\theta,\bar{\theta})$ in ${\mathbb R}^p\times {\mathbb R}^p$.
\item
[(ii)] $\lim_{\|\theta\|\rightarrow\infty}l_y(\theta)=-\infty$ where
$\|\theta\|$ is the standard Euclidean norm on ${\mathbb R}^p$.
\item [(iii)] $l_y(\theta) < \infty$ and $\Lambda_{\nabla^2 l_y(\theta)} <0$
  on every bounded $\theta$-set.
\item
[(iv)] for any $\bar{\theta}$ in ${\mathbb R}^p$,
$I_y(\bar{\theta},\theta) < \infty$ and $0
<\lambda_{\nabla_{01}^2I_y(\bar{\theta},\theta)} \leq
\Lambda_{\nabla_{01}^2I_y(\bar{\theta},\theta)}$ on every bounded
$\theta$-set.
\end{itemize}

\end{assumptions}

These assumptions ensure smoothness of $l_y(\theta)$ and
$I_y(\bar{\theta},\theta)$ and their first two derivatives in $\theta$.
Assumption \ref{ass}.iii also implies strong concavity of $l_y(\theta)$.
Assumption \ref{ass}.iv implies that $I_y(\bar{\theta},\theta)$ is strictly
convex and that the parameter $\theta$ is strongly identifiable in the family
of densities $k(x|y;\theta)$ (see proof of Lemma \ref{lemma:ident} below).
Note that the above assumptions are not the minimum possible set, e.g.  that
$l_y(\theta)$ and $I_y(\bar{\theta},\theta)$ are upper bounded follows from
continuity, Assumption \ref{ass}.ii and the property $I_y(\bar{\theta},\theta)
\geq I_y(\bar{\theta} ,\bar{\theta}) = 0$, respectively.

We first characterize the fixed points of the KPP algorithm. 

A result that will be used repeatedly in the sequel is that for any
$\bar{\theta} \in {\mathbb R}^p$
\begin{equation}
\nabla_{01}I_y(\bar{\theta},\bar{\theta}) =0 .
\label{simple}
\end{equation}
This follows immediately from the information inequality 
for the KL divergence \cite[Thm. 2.6.3]{Cover&Thomas:91}
$$I_y(\bar{\theta},\theta) \geq I_y(\bar{\theta},\bar{\theta}) =0,$$ 
so that, by smoothness
Assumption \ref{ass}.i, $I_y(\bar{\theta},\theta)$ has a stationary point at
$\theta=\bar{\theta}$.

\begin{proposition}
  Let the densities $g(y;\theta)$ and $k(x|y;\theta)$ be such that
  Assumptions \ref{ass} are satisfied. Then the fixed points
  of the recurrence in (\ref{iteration}) are maximizers of the log-likelihood
  function $l_y(\theta)$ for any relaxation sequence $\beta_k = \beta > 0$,
  $k=1,2,\ldots$.
\end{proposition}

\noindent{ \it Proof:}
Consider a fixed point $\theta^*$ of the recurrence relation
(\ref{iteration}) for $\beta_k=\beta=$ constant.  Then,
\begin{equation*}
\theta^*={\rm argmax}_{\theta \in {\mathbb R}^p} \left\{l_y(\theta)-
\beta I_y(\theta^*,\theta)\right\}.
\end{equation*}
As $l_y(\theta)$ and $I_y(\theta^*,\theta)$ are both smooth in $\theta$,
$\theta^*$ must be a stationary point
\begin{equation*}
0=\nabla l_y(\theta^*)-\beta \nabla_{01}I_y(\theta^*,\theta^*).
\end{equation*}
Thus, as by (\ref{simple}) $\nabla_{01}I_y(\theta^*,\theta^*) =0$,  
\begin{equation}
\label{statio}
0=\nabla l_y(\theta).
\end{equation}
Since $l_y(\theta)$ is 
strictly concave, we deduce that $\theta^*$ is a maximizer of $l_y(\theta)$.
\qed


The following will be useful. 
\begin{lemma}
\label{lemma:ident}
Let the conditional density $k(x|y;\theta)$ be such that
$I_y(\bar{\theta},\theta)$ satisfies Assumption \ref{ass}.iv. Then, given two
bounded sequences $\{\theta_1^k\}$ and $\{\theta_2^k\}$, $\lim_{k\rightarrow
  \infty} I_y(\theta_1^k,\theta_2^k)=0$ implies that $\lim_{k\rightarrow
  \infty} \|\theta_1^k-\theta_2^k\|=0$.
\end{lemma}

\noindent{ \it Proof:}
Let $\mathcal B$ be any bounded set containing both sequences
$\{\theta^k_1\}$ and $\{\theta_2^k\}$. Let $\lambda$ denote
the minimum 
\begin{equation}
\lambda=\min_{\theta,\bar{\theta}\in \mathcal B} 
\lambda_{\nabla_{01}^2 I_y(\bar{\theta},\theta)}
\end{equation}
Assumption \ref{ass}.iv implies that $\lambda > 0$. Furthermore,
invoking Taylor's theorem with remainder, $I_y(\bar{\theta},\theta)$ is
strictly convex in the sense that for any $k$
\begin{align}
I_y(\theta^k_1,\theta^k_2)\geq I_y(\theta^k_1,\theta^k_1)+
& \nabla I_y(\theta_1^k,\theta^k_1)^{\sf T}(\theta^k_1-\theta^k_2)\nonumber\\
& +\frac12 \lambda \|\theta^k_1-\theta^k_2\|^2.\nonumber
\end{align}
As $I_y(\theta^k_1,\theta^k_1)=0$ and $\nabla_{01}
I_y(\theta^k_1,\theta^k_1)=0$, recall (\ref{simple}), we obtain
\begin{equation}
I_y(\theta^k_1,\theta^k_2)\geq \frac{\lambda}2 \|\theta^k_1-\theta^k_2\|^2.\nonumber
\end{equation}
The desired result comes from passing to the limit $k \rightarrow \infty$.
\qed

Using these results, we easily obtain the following.
\begin{lemma}
\label{tough}
Let the densities $g(y;\theta)$ and $k(x|y;\theta)$ be such that Assumptions
\ref{ass} are satisfied. Then $\{\theta^k\}_{k\in \mathbb N}$ is bounded.
\end{lemma}

\noindent{ \it Proof:}
  Due to Proposition \ref{inc},  the sequence $\{l_y(\theta^k)\}$ is monotone
  increasing.  Therefore, assumption \ref{ass}.ii implies that $\{\theta^k\}$
  is bounded.
\qed

In the following lemma, we prove a result which is often called asymptotic
regularity \cite{Bauschke:1996}.

\begin{lemma}
\label{ignoble}
Let the densities $g(y;\theta)$ and $k(x|y;\theta)$ be such that
$l_y(\theta)$ and $I_y(\bar{\theta},\theta)$ satisfy Assumptions \ref{ass}.
Let the sequence of relaxation parameters $\{\beta_k\}_{k\in \mathbb N}$
satisfy $0< \liminf \beta_k \leq \limsup \beta_k < \infty$.  Then,
\begin{equation}
\lim_{k\rightarrow\infty} \|\theta^{k+1}-\theta^k\|=0.
\end{equation}
\end{lemma}

\noindent{ \it Proof:}
By Assumption \ref{ass}.iii and by Proposition \ref{inc}
$\{l_y(\theta^k)\}_{k\in \mathbb N}$ is bounded and monotone.  Since, by Lemma
\ref{tough}, $\{\theta^k\}_{k\in\mathbb N}$ is a bounded sequence 
$\{l_y(\theta^k)\}_{k\in \mathbb N}$
converges.  Therefore, $\lim_{k\rightarrow\infty}\left\{
l_y(\theta^{k+1})-l_y(\theta^k)\right\}=0$ which, from (\ref{gultov}),
implies 
%
that $\beta_k I_y(\theta^k,\theta^{k+1})$ vanishes when $k$ tends to
infinity. Since $\{\beta_k\}_{k\in \mathbb N}$ is bounded below by $\liminf
\beta_k>0$:
$\lim_{k\rightarrow\infty} I_y(\theta^k,\theta^{k+1})=0$. 
Therefore, Lemma \ref{lemma:ident} establishes the desired result.
\qed

We can now give a global convergence theorem.
\begin{theorem}
\label{Conv}
Let the sequence of relaxation parameters $\{\beta_k\}_{k\in \mathbb N}$ be
positive and converge to a limit $\beta^* \in [0,\infty)$.  Then the sequence
$\{\theta^k\}_{k\in \mathbb N}$ converges to the solution of the ML
estimation problem (\ref{ml}).
\end{theorem}

\noindent{ \it Proof:}
Since $\{\theta^k\}_{k\in \mathbb N}$ is bounded, one can extract a
convergent subsequence $\{\theta^{\sigma(k)}\}_{k\in \mathbb N}$ with limit
$\theta^*$. The defining recurrence (\ref{iteration}) implies that
\begin{equation}
\label{orugl}
\nabla l_y(\theta^{\sigma(k)+1})-\beta_{\sigma(k)}
\nabla_{01} I_y(\theta^{\sigma(k)},\theta^{\sigma(k)+1})=0.
\end{equation}
We now prove that $\theta^*$ is a stationary point of $l_y(\theta)$. 
Assume first that 
$\{\beta_k\}_{k\in \mathbb N}$ converges to zero, i.e. $\beta^*=0$. 
Due to Assumptions \ref{ass}.i, $\nabla l_y(\theta)$ is continuous in 
$\theta$. Hence, since $\nabla_{01} I_y(\bar{\theta},\theta)$ is 
bounded on bounded subsets, (\ref{orugl}) implies
\begin{equation}
\nabla l_y(\theta^*)=0.\nonumber
\nonumber
\end{equation}

Next, assume that $\beta^*>0$. In this case, Lemma \ref{ignoble} establishes
that
\begin{equation}
\lim_{k\rightarrow\infty} \|\theta^{k+1}-\theta^k\|=0.\nonumber
\end{equation}
Therefore, $\{\theta^{\sigma(k)+1}\}_{k\in \mathbb N}$ also tends to
$\theta^*$. Since $\nabla_{01} I_y(\bar{\theta},\theta)$ is continuous in
$(\bar{\theta}, \theta)$ equation (\ref{orugl}) gives at infinity
\begin{equation}
\label{orult}
\nabla l_y(\theta^*)-\beta^*
\nabla_{01} I_y(\theta^*,\theta^*)=0.\nonumber
\end{equation}
Finally, by (\ref{simple}), $\nabla_{01} I_y(\theta^*,\theta^*)=0$ and
\begin{equation}
\nabla l_y(\theta^*)=0.
\end{equation}

The proof is concluded as follows.  As, by Assumption \ref{ass}.iii,
$l_y(\theta)$ is concave, $\theta^*$ is a maximizer of $l_y(\theta)$ so that
$\theta^*$ solves the Maximum Likelihood estimation problem (\ref{ml}).
Furthermore, as positive definiteness of $\nabla^2 l_y$ implies that
$l_y(\theta)$ is in fact strictly concave, this maximizer is unique.  Hence,
$\{\theta^k\}$ has only one accumulation point and $\{\theta^k\}$ converges
to $\theta^*$ which ends the proof.
\qed

We now establish the main result concerning speed of convergence. Recall that
a sequence $\{ \theta^k\}$ is said to converge superlinearly to a limit
$\theta^*$ if:
\begin{equation}
\lim_{k\rightarrow\infty}\frac{\|\theta^{k+1}-\theta^*\|}
{\|\theta^k-\theta^*\|}=0,\label{eq:superl} .
\end{equation}

\begin{theorem}
  Assume that the sequence of positive relaxation parameters
  $\{\beta_k\}_{k\in \mathbb N}$ converges to zero. Then, the sequence
  $\{\theta^k\}_{k\in \mathbb N}$ converges superlinearly to the solution of
  the ML estimation problem (\ref{ml}).
\end{theorem}

\noindent{ \it Proof:}
Due to Theorem \ref{Conv}, the sequence $\{\theta^k\}$ converges to
the unique maximizer $\theta_{ML}$ of $l_y(\theta)$.  Assumption
\ref{ass}.i implies that the gradient mapping $\nabla_{\theta}
\bigl(l_y(\theta)-\beta_k I_y(\theta_{ML},\theta)\bigr)$ is continuously
differentiable. Hence, we have the following Taylor expansion
about $\theta_{ML}$.
\begin{align}
\nabla l_y(\theta)- & \beta_k \nabla_{01} I_y(\theta_{ML},\theta)=
\nabla l_y(\theta_{ML})\nonumber\\
& -\beta_k\nabla_{01} I_y(\theta_{ML},\theta_{ML})\nonumber\\
& +\nabla^2 l_y(\theta_{ML})(\theta-\theta_{ML})\label{pudics}\\
& -\beta_k \nabla_{01}^2 I_y(\theta_{ML},\theta_{ML})(\theta-\theta_{ML})\nonumber\\
& +R(\theta-\theta_{ML}),\nonumber
\end{align}
where the remainder satisfies
\begin{equation}
\lim_{\theta\rightarrow \theta_{ML}} \frac{\|R(\theta-\theta_{ML})\|}
{\|\theta-\theta_{ML}\|}=0.\nonumber
\end{equation}
Since $\theta_{ML}$ maximizes $l_y(\theta)$, $\nabla l_y(\theta_{ML})=0$.
Furthermore, by (\ref{simple}), $\nabla_{01} I_y(\theta_{ML},\theta_{ML})=0$.
Hence, (\ref{pudics}) can be simplified to
\begin{align}
\label{simplea}
\nabla l_y(\theta) &-\beta_k \nabla_{01} I_y(\theta_{ML},\theta) =
\nabla^2 l_y(\theta_{ML})(\theta-\theta_{ML})\nonumber\\
& -\beta_k \nabla_{01}^2 I_y(\theta_{ML},\theta_{ML})(\theta-\theta_{ML})+
R(\theta-\theta_{ML}).
\end{align}
From the defining relation (\ref{iteration}) the iterate
$\theta^{k+1}$ satisfies
\begin{equation}
\label{next}
\nabla l_y(\theta^{k+1})-\beta_k \nabla_{01} I_y(\theta^k,\theta^{k+1}) =0.
\end{equation}
So, taking $\theta=\theta^{k+1}$ in (\ref{simplea}) and using
(\ref{next}), we obtain
\begin{align}
& \beta_k\bigl(\nabla_{01} I_y(\theta^k,\theta^{k+1})-
\nabla_{01} I_y(\theta_{ML},\theta^{k+1}) \bigr)= \nonumber\\
& +\nabla^2 l_y(\theta_{ML})(\theta^{k+1}-\theta_{ML})-\beta_k 
\nabla_{01}^2 I_y(\theta_{ML},\theta_{ML})(\theta^{k+1}-\theta_{ML}) \nonumber\\
& + R(\theta^{k+1}-\theta_{ML}).\nonumber
\end{align}
Thus,
\begin{align}
\label{tentach}
& \|\beta_k\bigl(\nabla_{01} I_y(\theta^k,\theta^{k+1})-
\nabla_{01} I_y(\theta_{ML},\theta^{k+1}) \bigr)-R(\theta^{k+1}-\theta_{ML})\|=\nonumber\\
& \|\nabla^2 l_y(\theta_{ML})(\theta^{k+1}-\theta_{ML})-\beta_k 
\nabla_{01}^2 I_y(\theta_{ML},\theta_{ML})(\theta^{k+1}-\theta_{ML})\|.
\end{align}

On the other hand, one deduces from Assumptions \ref{ass} (i) that
$\nabla_{01} I_y(\bar{\theta},\theta)$ is locally Lipschitz in the variables
$\theta$ and $\bar{\theta}$.  Then, since, $\{\theta^k\}$ is bounded, there
exists a bounded set $\mathcal B$ containing $\{\theta^k\}$ and a finite
constant $L$ such that for all $\theta$, $\theta^{\prime}$, $\bar{\theta}$
and $\bar{\theta}^{\prime}$ in $\mathcal B$,
\begin{equation}
\|\nabla_{01} I_y(\bar{\theta},\theta)-\nabla_{01} I_y(
\bar{\theta}^{\prime},\theta^{\prime})\| \leq L \bigl(\|\theta-\theta^{\prime}\|^2+
\|\bar{\theta}-\bar{\theta}^{\prime}\|^2\bigr)^{\frac12}.\nonumber
\end{equation}
Using the triangle inequality and this last result,  
(\ref{tentach}) asserts that for any $\theta \in \mathcal B$
\begin{align}
\label{tenkash}
\beta_k L \|\theta^{k}-\theta_{ML}\| & +\|R(\theta^{k+1}-\theta_{ML})\|
\geq \|\bigl(\nabla^2 l_y(\theta_{ML})\nonumber \\
& -\beta_k \nabla_{01}^2
I_y(\theta_{ML},\theta_{ML})\bigr)(\theta^{k+1}-\theta_{ML})\| .
\end{align}
Now, consider again the bounded set $\mathcal B$ containing $\{\theta^k\}$. 
Let $\lambda_{l_y}$ and $\lambda_I$ denote the 
minima
\begin{equation}
\lambda_{l_y}=\min_{\theta \in \mathcal B} 
\left\{-\lambda_{\nabla^2 l_y(\theta)}\right\}
\nonumber
\end{equation}
\begin{equation}
\lambda_{I}=\min_{\theta,\bar{\theta}\in \mathcal B} 
\left\{\lambda_{\nabla_{01}^2 I_y(\bar{\theta},\theta)}\right\}.
\nonumber
\end{equation}
Since for any symmetric matrix $H$, $x^T H x / \|x\|^2$ is lower
bounded by the minimum eigenvalue of $H$, we have immediately that
\begin{align}
& \|\bigl(-\nabla^2 l_y(\theta_{ML})+
\beta_k \nabla_{01}^2 I_y(\theta_{ML},\theta_{ML})\bigr)
(\theta^{k+1}-\theta_{ML})\|^2 \nonumber\\ 
& \geq \bigl(\lambda_{l_y}+\beta_k
\lambda_I\bigr)^2\|\theta^{k+1}-\theta_{ML}\|^2. \label{eq:Rayliegh}
\end{align}
By Assumptions \ref{ass}.iii and \ref{ass}.iv, $\lambda_{l_y}+\beta_k
\lambda_I > 0$ and, after substitution of (\ref{eq:Rayliegh}) into
(\ref{tenkash}), we obtain
\begin{align}
\beta_k L \|\theta^{k}-\theta_{ML}\| &
+\|R(\theta^{k+1}-\theta_{ML})\| \geq \nonumber \\ &
\bigl(\lambda_{l_y}+\beta_k
\lambda_I\bigr)\|\theta^{k+1}-\theta_{ML}\|,
\label{eq:ineq2}
\end{align}
for all $\theta \in \mathcal B$.
Therefore,  collecting terms  in (\ref{eq:ineq2})
\begin{equation}
\label{oprute}
\beta_k L \geq 
\left(\lambda_{l_y}+\beta_k\lambda_I-\frac{
\|R(\theta^{k+1}-\theta_{ML})\|}{\|\theta^{k+1}-\theta_{ML}\|}\right)
\frac{\|\theta^{k+1}-\theta_{ML}\|}{\|\theta^k-\theta_{ML}\|}.
\end{equation}
Now, recall that $\{\theta^k\}$ is convergent. Thus,
$\lim_{k\rightarrow\infty}\|\theta^k-\theta_{ML}\|=0$ and subsequently,
$\lim_{k\rightarrow\infty}
\frac{\|R(\theta^{k+1}-\theta_{ML})\|}{\|\theta^{k+1}-\theta_{ML}\|}=0$ due
to the definition of the remainder $R$. 
%
Finally, as $\beta_k$ converges to zero, $L$ is bounded and
$\lambda_{l_y}>0$, equation (\ref{oprute}) gives (\ref{eq:superl}) with
$\theta^* = \theta_{ML}$ and the proof of superlinear convergence is
completed.
\qed

\section{Second order Approximations and Trust Region techniques}
\label{approx}

The maximization in the KPP recursion (\ref{iteration}) will not generally
yield an explicit exact recursion in $\theta^k$ and $\theta^{k+1}$. Thus
implementation of the KPP algorithm methods may require line search or
one-step-late approximations similar to those used for the M-step of the
non-explicit penalized EM maximum likelihood algorithm \cite{Green:JRSS90}.
In this section, we discuss an alternative which uses second order function
approximations and preserves the convergence properties of KPP established in
the previous section. This second order scheme is related to the well-known
Trust Region technique for iterative optimization introduced by Mor\'e
\cite{More:83}.

\subsection{Approximate models}

In order to obtain computable iterations, the following second order
approximations of $l_y(\theta)$ and $I_y(\theta^k,\theta)$ are introduced
\begin{eqnarray}
\hat{l}_y(\theta)&=&l_y(\theta^k)+  \nabla l_y(\theta^k)^{\sf T}
(\theta-\theta^k)+\nonumber\\ &&
 \frac12 (\theta-\theta^k)^{\sf T} H_k
(\theta-\theta^k). \nonumber
\label{eq:quad}
\end{eqnarray}
and
\begin{equation}
\hat{I}_y(\theta,\theta^k)=
\frac12 (\theta-\theta^k)^{\sf T}\nabla_{01}^2 I_k(\theta-\theta^k).
\nonumber
\end{equation}
In the following, we adopt the simple notation $g_k=\nabla l_y(\theta^k)$ (a
column vector). A natural choice for $H_k$ and $I_k$ is of course
\begin{equation}
\label{hess1}
H_k=\nabla^2 l_y(\theta^k) \nonumber
\end{equation}
and
\begin{equation}
\label{hess2}
I_k=\nabla_{01}^2 I_y(\theta^k,\theta^k).\nonumber
\end{equation}
The approximate KPP algorithm is defined as
\begin{align}
\label{aKprox}
\theta^{k+1}={\rm argmax}_{\theta \in {\mathbb R}^p} &
\bigl\{ 
l_y(\theta^k)+g_k(\theta-\theta^k) \nonumber\\
& +\frac12(\theta-\theta^k)^{\sf T} H_k (\theta-\theta^k) \\
& -\frac{\beta_k}{2} (\theta-\theta^k)^{\sf T}I_k (\theta-\theta^k)
\bigr\}\nonumber
\end{align}

At this point it is important to make several comments.  Notice first that
for $\beta_k=0$, $k=1,2,\ldots$, and $H_k=\nabla^2 l_y(\theta^k)$, the
approximate step (\ref{aKprox}) is equivalent to a Newton step. It is well
known that Newton's method, also known as Fisher scoring, has superlinear
asymptotic convergence rate but may diverge if not properly initialized.
Therefore, at least for small values of the relaxation parameter $\beta_k$,
the approximate PPA algorithm may fail to converge for reasons analogous in
Newton's method \cite{Ortega&Rheinboldt:70}.  On the other hand, for $\beta_k
>0$ the term $-\frac{\beta_k}{2} (\theta-\theta^k)^{\sf T}I_k
(\theta-\theta^k)$ penalizes the distance of the next iterate $\theta^{k+1}$
to the current iterate $\theta^k$. Hence, we can interpret this term as a
regularization or relaxation which stabilizes the possibly divergent Newton
algorithm without sacrificing its superlinear asymptotic convergence rate.
By appropriate choice of $\{\beta_k\}$ the iterate $\theta^{k+1}$ can be
forced to remain in a region around $\theta^k$ over which the quadratic model
$\hat{l}_y(\theta)$ is accurate \cite{More:83}\cite{Bonnans:1997}.
%

In many cases a quadratic approximation of a single one of the two terms
$l_y(\theta)$ or $I_y(\theta^k,\theta)$ is sufficient to obtain a closed form
for the maximum in the KPP recursion (\ref{iteration}).  Naturally, when
    feasible, such a reduced approximation is preferable to the approximation
    of both terms discussed above.  For concreteness, in the sequel, although
    our results hold for the reduced approximation also, we only prove
    convergence for the proximal point algorithm implemented with the full
    two-term approximation.
    
    Finally, note that (\ref{aKprox}) is quadratic in $\theta$ and the
    minimization problem clearly reduces to solving a linear system of
    equations. For $\theta$ of moderate dimension, these equations can be
    efficiently solved using conjugate gradient techniques \cite{Nocedal:1999}.
    However, when the vector $\theta$ in (\ref{aKprox}) is of large
    dimension, as frequently occurs in inverse problems, limited memory BFGS
    quasi-Newton schemes for updating $H_k-\beta_k I_k$ may be
    computationally much more efficient, see for example \cite{Nocedal:1999},
    \cite{Nocedal:1980}, \cite{Liu:1989}, \cite{Gilbert:1989} and
    \cite{Fletcher:1991}.  
%

\subsection{Trust Region Update Strategy}

The Trust Region strategy proceeds as follows. The model $\hat{l}_y(\theta)$
is maximized in a ball $B(\theta^k,\delta)= \bigl\{\|\theta-\theta^k\|_{I_k}
\leq \delta\bigr\}$ centered at $\theta^k$ where $\delta$ is a proximity
control parameter which may depend on $k$, and where $\|a\|_{I_k}= a^{\sf
  T}I_k a$ is a norm; well defined due to positive definiteness of $I_k$
(Assumption \ref{ass}.iv).  Given an iterate $\theta^k$ consider a candidate
$\theta^\delta$ for $\theta^{k+1}$ defined as the solution to the constrained
optimization problem
\begin{equation}
\theta^{\delta}={\rm argmax}_{\theta \in {\mathbb R}^p}\hat{l}_y(\theta)\nonumber
\end{equation}
subject to
\begin{equation}
\|\theta-\theta^k\|_{I_k}\leq \delta. \label{eq:constraint}
\end{equation}
By duality theory of constrained optimization \cite{Hiriart&Lemarechal:93},
and the fact that $\hat{l}_y(\theta)$ is strictly concave, this
problem is equivalent to the unconstrained optimization
\begin{equation}
\label{dual}
\theta^\delta(\beta)={\rm argmin}_{\theta\in {\mathbb R}^p}
L(\theta,\beta).
\end{equation}
where
\begin{equation}
L(\theta,\beta)=-\hat{l}_y(\theta)
+\frac\beta 2\bigl(\|\theta-\theta^k\|_{I_k}^2-\delta^2\bigr).\nonumber
\end{equation}
and $\beta$ is a Lagrange multiplier selected to meet the
constraint (\ref{eq:constraint}) with equality:
$\|\theta^\delta(\beta) - \theta\|_{I_k} = \delta$.

We conclude that the Trust Region candidate $\theta^\delta$ is identical to
the approximate KPP iterate (\ref{aKprox}) with relaxation parameter
$\beta$ chosen according to constraint (\ref{eq:constraint}).  This relation
also provides a rational rule for computing the relaxation parameter $\beta$.

\subsection{Implementation}
\label{Impl}
The parameter $\delta$ is said to be safe if $\theta^\delta$ produces an
acceptable increase in the original objective $l_y$.  An iteration of the
Trust Region method consists of two principal steps

{\em Rule 1}. Determine whether $\delta$ is safe or not. If $\delta$
is safe, set $\delta_k=\delta$ and take an approximate Kullback
proximal step $\theta^{k+1}=\theta^\delta$. Otherwise, take a {\it
null step} $\theta^{k+1}=\theta^k$.

{\em Rule 2}. Update $\delta$ depending on the result of {\it Rule 1}.

Rule 1 can be implemented by 
comparing the increase in the original log-likelihood $l_y$ to a
fraction $m$ of the expected increase predicted by the approximate
model $\hat{l}_y(\theta)$.  Specifically, the Trust Region parameter $\delta$
is accepted if
\begin{equation}
l_y(\theta^\delta)-l_y(\theta^k)\geq m 
\bigl(\hat{l}_y(\theta^\delta)-\hat{l}_y(\theta^k) \bigr).
\end{equation}
Rule 2 can be implemented as follows. If $\delta$ was accepted by Rule
1, $\delta$ is increased at the next iteration in order to extend the
region of validity of the model $\hat{l}_y(\theta)$. If $\delta$ was
rejected, the region must be tightened and $\delta$ is decreased at
the next iteration.

The Trust Region strategy implemented here is essentially the same as that 
proposed by Mor\'e 
\cite{More:83}. 
\begin{algorithm}
\label{alg}
Step 0. (Initialization) Set $\theta^0 \in {\mathbb R}^p$, $\delta_0>0$ and
the ``curve search'' parameters $m$, $m^{\prime}$ with $0<m<m^{\prime}<1$. 

Step 1. With $\hat{l}_y(\theta)$ the quadratic approximation (\ref{eq:quad}),
solve 
\begin{equation}
\theta^{\delta_k}={\rm argmax}_{\theta \in {\mathbb R}^p}\hat{l}_y(\theta)\nonumber
\end{equation}
subject to
\begin{equation}
\|\theta-\theta^k\|_{I_k}\leq \delta_k. \nonumber
\end{equation}

Step 2. If $l_y(\theta^{\delta_k})-l_y(\theta^k)\geq m 
\bigl(\hat{l}_y(\theta^{\delta_k})-\hat{l}_y(\theta^k) \bigr)$ then set 
$\theta^{k+1}=\theta^{\delta_k}$. Otherwise, set $\theta^{k+1}=\theta^k$.

Step 3. Set $k=k+1$. Update the model $\hat{l}_y(\theta^k)$. Update
$\delta_k$ using Procedure \ref{upd}.

Step 4. Go to Step 1.
\end{algorithm}

The procedure for updating $\delta_k$ is given below. 
\begin{procedure}
\label{upd}
Step 0. (Initialization) Set $\gamma_1$ and $\gamma_2$ such that
$\gamma_1<1<\gamma_2$. 

Step 1. If $l_y(\theta^{\delta_k})-l_y(\theta^k)\leq m 
\bigl(\hat{l}_y(\theta^{\delta_k})-\hat{l}_y(\theta^k) \bigr)$ then
take $\delta_{k+1} \in (0,\gamma_1 \delta_k)$. 

Step 2. If $l_y(\theta^{\delta_k})-l_y(\theta^k)\leq m^{\prime} 
\bigl(\hat{l}_y(\theta^{\delta_k})-\hat{l}_y(\theta^k) \bigr)$ then
take $\delta_{k+1} \in (\gamma_1 \delta_k,\delta_k)$.

Step 3. If $l_y(\theta^{\delta_k})-l_y(\theta^k)\geq m^{\prime} 
\bigl(\hat{l}_y(\theta^{\delta_k})-\hat{l}_y(\theta^k) \bigr)$ then
take $\delta_{k+1} \in (\delta_k,\gamma_2\delta_k)$.
\end{procedure}

The Trust Region algorithm satisfies the following convergence theorem 

\begin{theorem}
\label{thm:more}
Let $g(y;\theta)$ and $k(x|y;\theta)$ be such that Assumptions 1 are
satisfied. Then, $\{\theta^k\}$ generated by Algorithm \ref{alg}
converges to the maximizer $\theta_{ML}$ of the log-likelihood
$l_y(\theta)$ and satisfies the monotone likelihood property $l_y(\theta^{k+1}) \geq l_y(\theta^k)$. 
If in addition, the sequence of Lagrange multipliers
$\{\beta_k\}$ tends towards zero, $\{\theta^k\}$ converges
superlinearly. 
\end{theorem}

The proof of Theorem \ref{thm:more} is omitted since it is standard in the
analysis of Trust Region methods; see \cite{More:83,Nocedal:1999}.
Superlinear convergence for the case that $\lim_{k\rightarrow \infty} \beta_k
=0$ follows from the Dennis and Mor\'e criterion \cite[Theorem
3.11]{Bonnans:1997}.

\subsection{Discussion}

The convergence results of Theorems 1 and 2 apply to any class of objective
functions which satisfy the Assumptions \ref{ass}. For instance, the analysis
directly applies to the penalized maximum likelihood (or posterior
likelihood) objective function $l^{'}_y(\theta)=l_y(\theta)+p(\theta)$ when
the ML penalty function (prior) $p(\theta)$ is quadratic and non-negative of
the form $p(\theta) = (\theta-\theta_o)^T R (\theta-\theta_o)$, where $R$ is
a non-negative definite matrix.  

The convergence Theorems 1 and 2 make use of concavity of $l_y(\theta)$ and
convexity of $I_y(\bar{\theta},\theta)$ via Assumptions \ref{ass}.iii and
\ref{ass}.iv. However, for smooth non-convex functions an analogous local
superlinear convergence result can be established under somewhat stronger
assumptions similar to those used in \cite{Hero&Fessler:SS95}.
  Likewise the Trust Region framework can also be applied to nonconvex objective
  functions.  In this case, 
  global convergence to a local maximizer of $l_y(\theta)$ can be established
  under Assumptions \ref{ass}.i, \ref{ass}.ii and \ref{ass}.iv following the
  proof technique of \cite{More:83}.

\section{Application to Poisson data}
\label{sec:numerical}

In this section, we illustrate the application of Algorithm \ref{alg} for a
maximum likelihood estimation problem in a Poisson inverse problem arising in
radiography, thermionic emission processes, photo-detection, and positron
emission tomography (PET).

\subsection{The Poisson Inverse Problem}

The objective is to estimate the intensity vector
$\theta=[\theta_1,\ldots,\theta_p]^T$ governing the number of gamma-ray
emissions $N=[N_1,\ldots,N_p]^T$ over an imaging volume of $p$ pixels. The
estimate of $\theta$ must be based on a vector of $m$ observed projections of
$N$ denoted $Y=[Y_1,\ldots,Y_m]^T$. The components $N_i$ of $N$ are
independent Poisson distributed with rate parameters $\theta_i$, and the
components $Y_j$ of $Y$ are independent Poisson distributed with rate
parameters $\sum_{i=1}^p P_{ji}\theta_i$, where $P_{ji}$ is the transition
probability; the probability that an emission from pixel $i$ is detected at
detector module $j$. The standard choice of complete data $X$, introduced by
Shepp and Vardi \cite{Shepp&Vardi:MI82}, for the EM algorithm is the set
$\{N_{ji}\}_{1\leq j \leq m,\;\;1\leq i \leq p}$, where $N_{ji}$ denotes the
number of emissions in pixel $i$ which are detected at detector $j$. The
corresponding many-to-one mapping $h(X)=Y$ in the EM algorithm is
\begin{equation}
Y_j=\sum_{i=1}^p N_{ji},\;\;\; 1\leq j \leq m.
\end{equation} 
It is also well known \cite{Shepp&Vardi:MI82} that the likelihood function is
given by
\begin{equation}
\log g(y;\theta)=\sum_{j=1}^m \Big(\sum_{i=1}^pP_{ji}\theta_i\Big)
-y_j \log \Big(\sum_{i=1}^pP_{ji}\theta_i\Big)+\log y_j!
\end{equation}
and that the expectation step of the EM algorithm is (see \cite{Green:JRSS90})
\begin{equation}\label{Exp}
Q(\theta,\bar{\theta}) = {\sf E}[\log f(x;\theta)\mid y;\bar{\theta}]=
\end{equation}
\vspace{-.5cm}
\begin{equation}
\sum_{j=1}^m \sum_{i=1}^p \Big( \frac{y_j P_{ji}\bar{\theta}_i}{\sum_{i=1}^p 
P_{ji} \bar{\theta}_i} \log (P_{ji}\theta_i )-P_{ji} \theta_i \Big).\nonumber
\end{equation}

Let us make the following additional assumptions: 
\begin{itemize}
\item
the solution(s)  of the Poisson inverse problem is (are) positive
\item
the level set 
\begin{equation}
\mathcal L=\{ \theta\in \mathbb R^n \mid l_y(\theta)\geq l_y(\theta^1) \}
\end{equation}
is bounded and included in the positive orthant. 
\end{itemize}
Then, since $l_y$ is continuous, $\mathcal L$ is compact. Due to the
monotonicity property of $\{\theta^k\}$, we thus deduce that for all $k$,
$\theta_i^k\geq \gamma$ for some $\gamma>0$. Then, the likelihood function
and the regularization function are both twice continuously differentiable on
the closure of $\{\theta^k\}$ and the theory developed in this paper applies.
These assumptions are very close in spirit to the assumptions in Hero and
Fessler \cite{Hero&Fessler:SS95}, except that we do not require the maximizer
to be unique.  The study of KPP without these assumptions requires further
analysis and is addressed in \cite{Chretien&Hero:SIAM98}.

\subsection{Simulation results}

\begin{figure}
\begin{center}
\makebox[8 cm][l]{
\vbox to 7 cm{
\vfill
\includegraphics{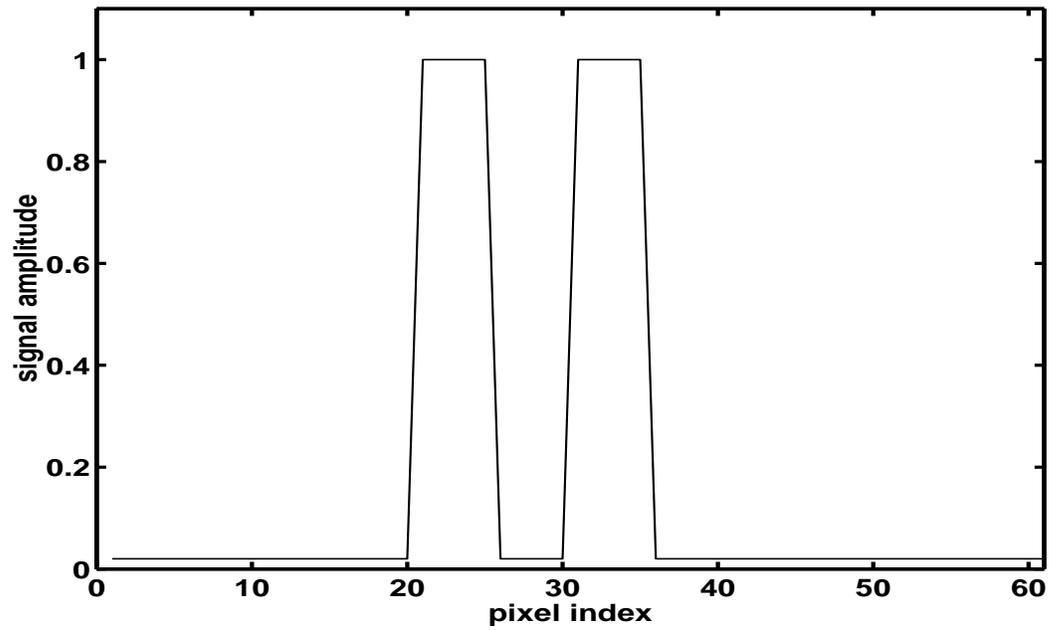}
}}
\vskip 0.2in
\caption{Two rail phantom for 1D deblurring example.}
\label{f1}
\end{center}
\end{figure}

\begin{figure}
\begin{center}
  \makebox[8 cm][l]{ \vbox to 7 cm{ \vfill \includegraphics{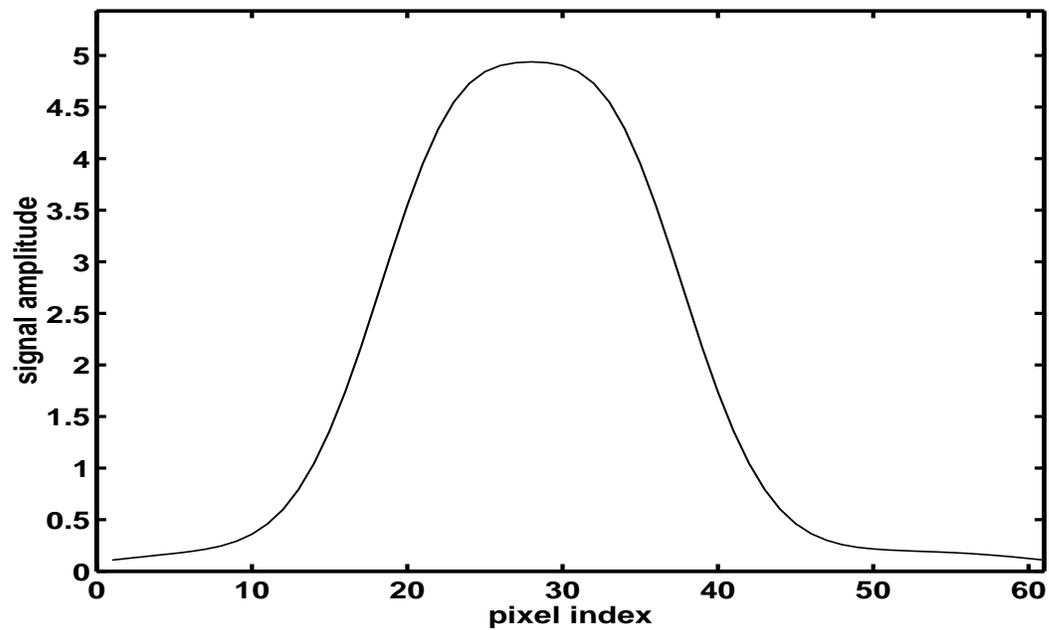} }}
\vskip 0.2in
\caption{Blurred two level phantom. Blurring kernel is Gaussian with standard
  width approximately equal to rail separation distance in phantom. An
  additive randoms noise of 0.3 was added.}
\label{f2}
\end{center}
\end{figure}

For illustration we performed numerical optimization for a simple one
dimensional deblurring example under the Poisson noise model of the previous
section.  This example easily generalizes to more general 2 and 3 dimensional
Poisson deblurring, tomographic reconstruction, and other imaging
applications. The true source $\theta$ is a two rail phantom shown in Figure
\ref{f1}.  The blurring kernel is a Gaussian function yielding the blurred
phantom shown in Figure \ref{f2}. We implemented both EM and KPP with Trust
Region update strategy for deblurring Fig. \ref{f2} when the set of ideal
blurred data $Y_i = \sum_{j=1}^N P_{ij} \theta_j$ is available without
Poisson noise.
In this simple noiseless case the ML solution is equal to the true source
$\theta$ which is everywhere positive.  Treatment of this noiseless case
allows us to investigate the behavior of the algorithms in the asymptotic
high count rate regime. More extensive simulations with Poisson noise will be
presented elsewhere.

\begin{figure}
\begin{center}
\makebox[8 cm][l]{
\vbox to 7 cm{
\vfill
\includegraphics{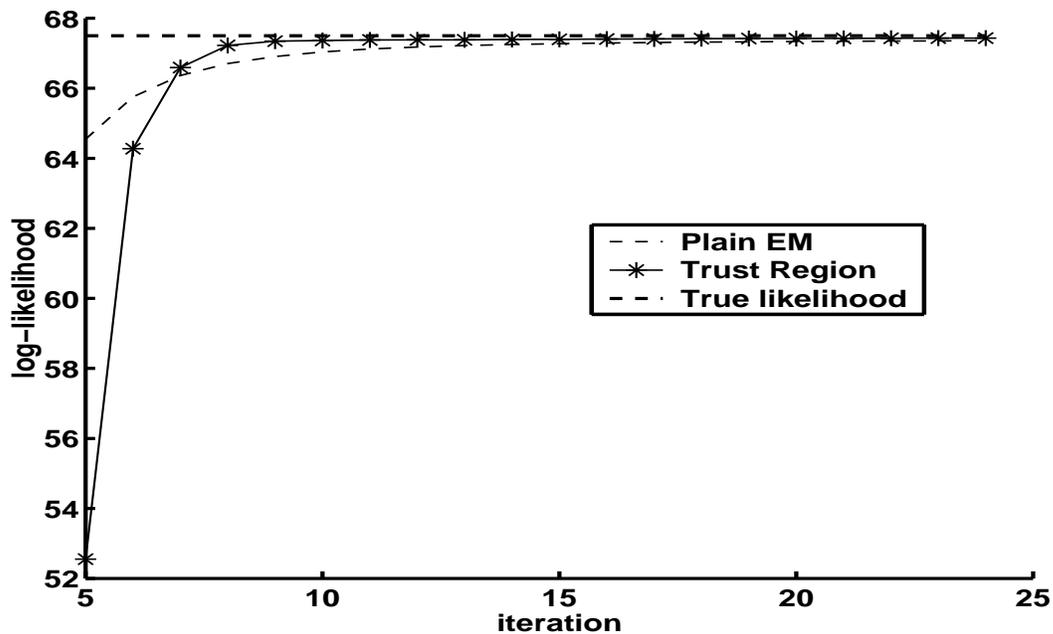}
}}
\vskip 0.2in
\caption{Snapshot of log--Likelihood vs iteration for plain 
  EM and KPP EM algorithm.  Plain EM initially produces greater increases in
  likelihood function but is overtaken by KPP EM at 7 iterations and
  thereafter.}
\label{f3}
\end{center}
\end{figure}

\begin{figure}
\begin{center}
\makebox[8 cm][l]{
\vbox to 7 cm{
\vfill
\includegraphics{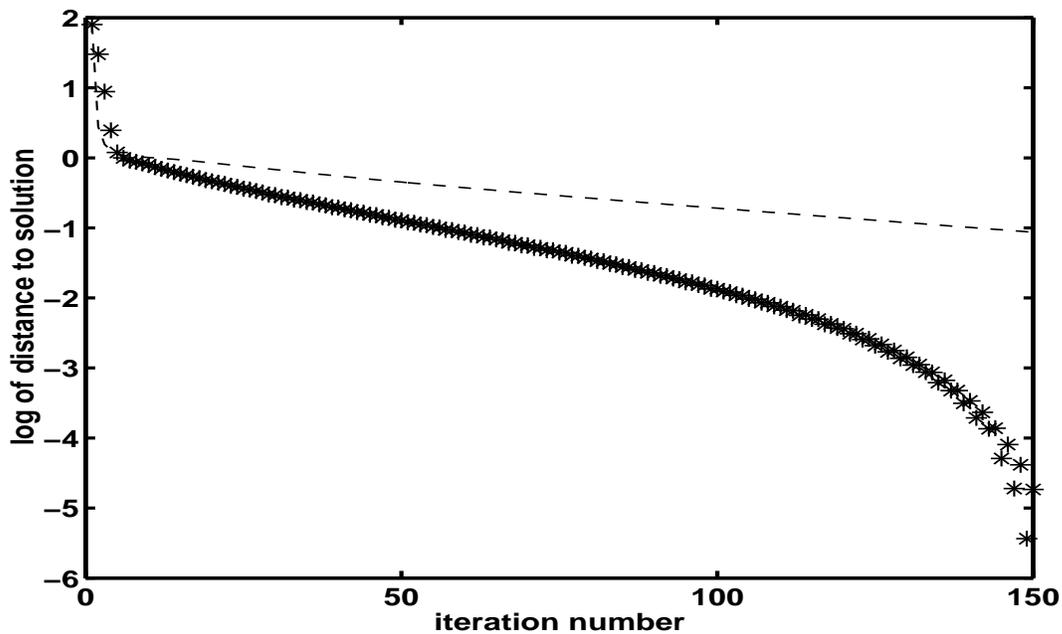}
}}
\vskip 0.2in
\caption{The sequence $\log \|\theta_k-\theta^*\|$ vs iteration for plain EM
  and  KPP EM algorithms.  Here $\theta^*$ is limiting value for each
  of the algorithms. Note the superlinear convergence of KPP. }
\label{f4}
\end{center}
\end{figure}

The numerical results shown in Fig. \ref{f3} indicate that the Trust Region
implementation of the KPP algorithm enjoys significantly faster convergence
towards the optimum than does EM.  
For these simulations the Trust Region technique was implemented in the
standard manner 
where the trust region size sequence $\delta_k$ in Algorithm 1 is determined
implicitly by the $\beta_k$ update rule: $\beta_{k+1} = 1.6 \beta_{k}$
($\delta_{k}$ is decreased) and otherwise $\beta_{k+1} = 0.5 \beta_{k}$
($\delta_{k}$ is increased).
The results shown in Fig.  \ref{f4} validate the theoretical superlinear
convergence of the Trust Region iterates as contrasted with the linear
convergence rate of the EM iterates.  Figure \ref{fig5.eps} shows the
reconstructed profile and demonstrates that the Trust Region updated KPP
technique achieves better reconstruction of the original phantom for a fixed
number of iterations.  Finally, Figure \ref{fig6.eps} shows the iterates for
the reconstructed phantom, plotted as a function of iteration on the
horizontal axis and as a function of grey level on the vertical axis.
Observe that the KPP achieves more rapid separation of the two components in
the phantom than does standard EM.

\section{Conclusions}

The main contributions of this paper are the following. First, we introduced
a general class of iterative methods for ML estimation based on
Kullback-Liebler relaxation of the proximal point strategy. Next, we proved
that the EM algorithm belongs to the proposed class, thus providing a new and
useful interpretation of the EM approach for ML estimation.  Finally, we
showed that Kullback proximal point methods enjoy global convergence and even
superlinear convergence for sequences of positive relaxation parameters that
converge to zero. Implementation issues were also discussed and we proposed
second order schemes for the case where the maximization step is hard to
obtain in closed form. We addressed Trust Region methodologies for the
updating of the relaxation parameters.  Computational experiments indicated
that the approximate second order KPP is stable and verifies the superlinear
convergence property as was predicted by our analysis.


\newpage

St\'ephane Chr\'etien - Biosketch

St\'ephane Chr\'etien was born in Rennes, France in 1969. He received the
B.S. and the Ph.D in Electrical Engineering from Universit\'e Paris Sud-Orsay
in 1992 and 1996 respectively. He then hold a postdoctoral position at the
EECS department of the University of Michigan, Ann Arbor and a research
position at INRIA Rh\^one-Alpes, France. He is now with the Service de
Math\'ematiques de la Gestion at the Universit\'e Libre de Bruxelles,
Belgium. His current research interests are in statistical estimation and
computational optimization with applications to image reconstruction and
urban traffic modelling and control.  

Alfred Hero - Biosketch 

Alfred O. Hero III, was born in Boston, MA. in 1955. He received the B.S. in
Electrical Engineering (summa cum laude) from Boston University (1980) and
the Ph.D from Princeton University (1984), both in Electrical
Engineering. While at Princeton he held the G.V.N. Lothrop Fellowship in
Engineering. Since 1984 he has been with the Dept. of Electrical
Engineering and Computer Science at the University of Michigan, Ann Arbor,
where he is currently Professor and Director of the Communications and Signal
Processing Laboratory. He has held positions of Visiting Scientist at
M.I.T. Lincoln Laboratory (1987 - 1989), Visiting Professor at Ecole
Nationale des Techniques Avancees (ENSTA), Ecole Superieure d'Electricite,
Paris (1990), Ecole Normale Sup\'erieure de Lyon (1999), and Ecole Nationale
Sup\'erieure des T\'el\'ecommunications, Paris (1999), William Clay Ford
Fellow at Ford Motor Company (1993).  His current research interests are in
the area of estimation and detection, statistical communications, signal
processing, and image processing. Alfred Hero is a Fellow of the Institute of
Electrical and Electronics Engineers (IEEE), a member of Tau Beta Pi, the
American Statistical Association (ASA), and Commission C of the International
Union of Radio Science (URSI). He received the 1998 IEEE Signal Processing
Society Meritorious Service Award, the 1998 IEEE Signal Processing Society
Best Paper Award, and the IEEE Third Millenium Medal.

He has served as Associate Editor for the IEEE Transactions on Information
Theory. He was also Chairman of the Statistical Signal and Array Processing
(SSAP) Technical Committee of the IEEE Signal Processing Society. He served
as treasurer of the Conference Board of the IEEE Signal Processing
Society. He was Chairman for Publicity for the 1986 IEEE International
Symposium on Information Theory (Ann Arbor, MI). He was General Chairman for
the 1995 IEEE International Conference on Acoustics, Speech, and Signal
Processing (Detroit, MI). He was co-chair for the 1999 IEEE Information
Theory Workshop on Detection, Estimation, Classification and Filtering (Santa
Fe, NM) and the 1999 IEEE Workshop on Higher Order Statistics (Caesaria,
Israel). He is currently a member of the Signal Processing Theory and Methods
(SPTM) Technical Committee and Vice President (Finances) of the IEEE Signal
Processing Society. He is also currently Chair of Commission C (Signals and
Systems) of the US delegation of the International Union of Radio Science
(URSI).

\newpage

\newpage

\newpage 

\newpage

\newpage

\newcommand{\myepsfig}[4]{%
\begin{figure}[htbp]%
\centering%
\leavevmode%
\epsfxsize=#3%
\epsffile{#1/#2}%
\caption{\it #4}%
\label{#2}%
\end{figure}%
}

\myepsfig{.}{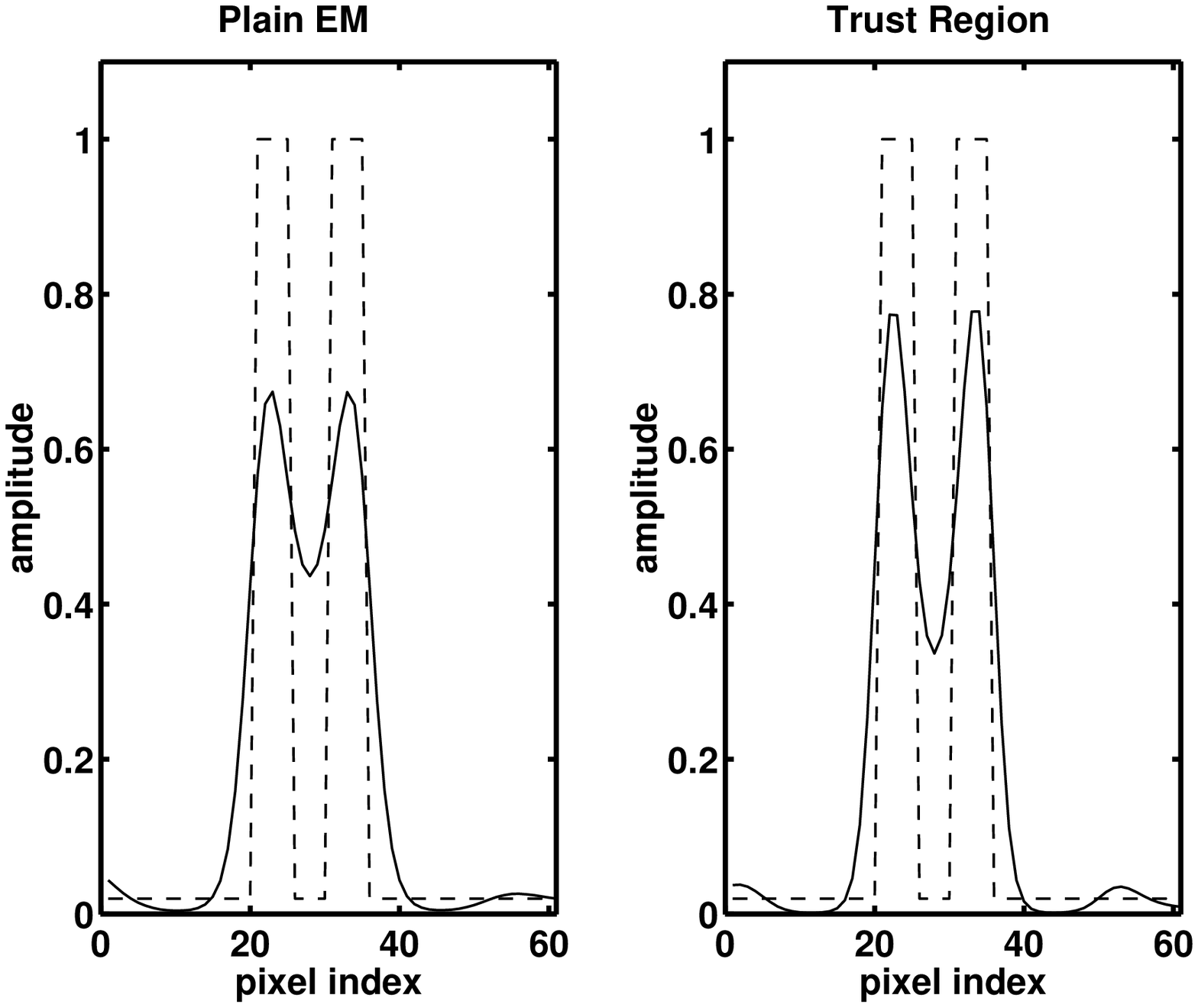}{5in}{Reconstructed images after 150 iterations of plain
  EM and KPP EM algorithms.}

\newpage

\myepsfig{.}{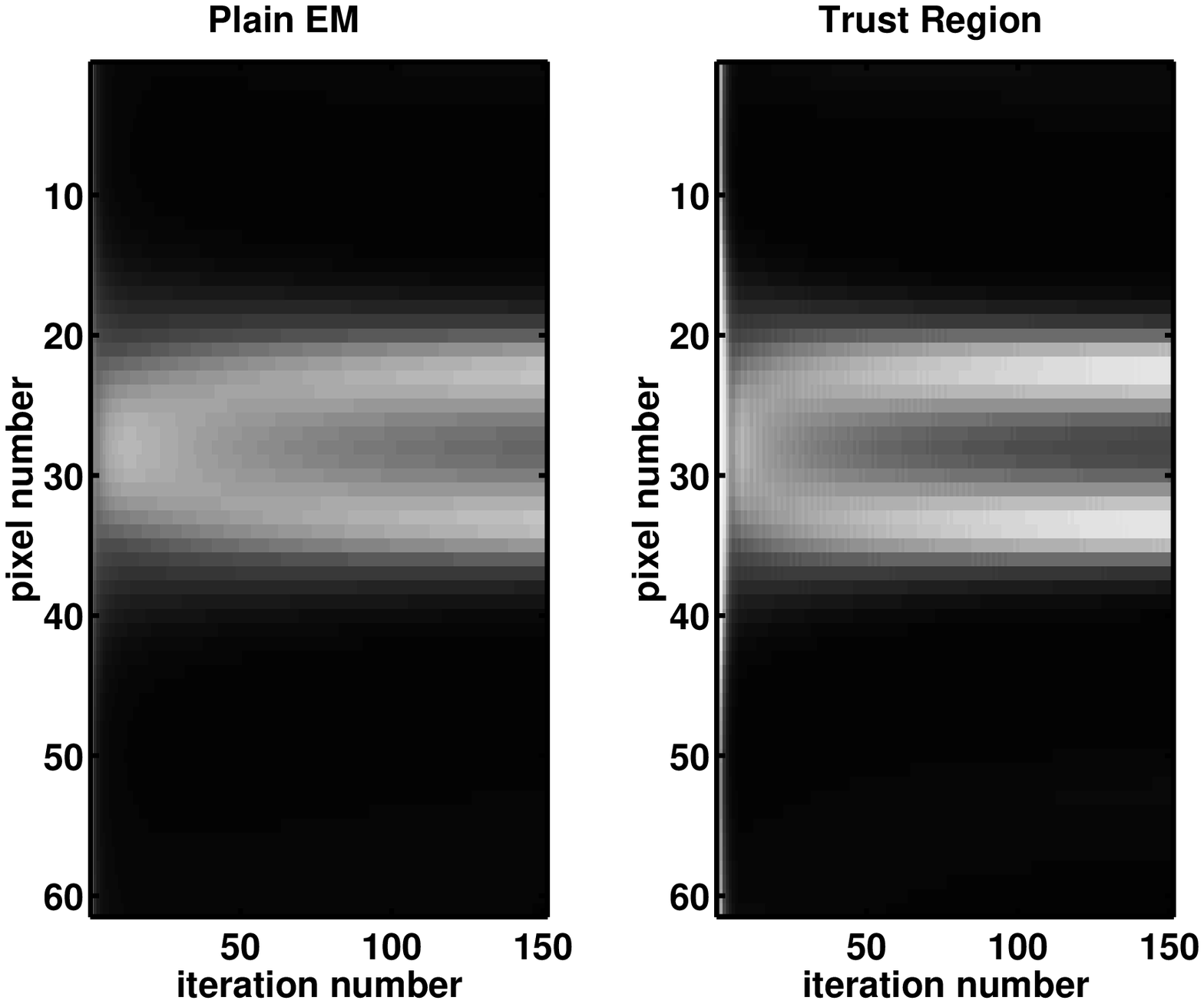}{5in}{Evolution of the reconstructed source vs iteration
  for plain EM and KPP EM.}

\end{document}